\documentclass[twocolumn]{aastex61}
\pdfoutput=1 
\usepackage{upgreek}

\def \frb {FRB\,121102}
\def \halpha {\ensuremath{\mathrm{H\alpha}}}
\def \hbeta {\ensuremath{\mathrm{H\beta}}}

\newcommand{\be}{\begin{eqnarray}}
\newcommand{\ee}{\end{eqnarray}}

\newcommand{\Halpha}{\rm H\alpha}
\newcommand{\EM}{{\rm EM}}
\newcommand{\DM}{{\rm DM}}

\newcommand{\SFR}{\ensuremath{\mathrm{SFR}}}

\newcommand{\DMhatFRB}{\rm \widehat{\DM}({\rm FRB})}

\newcommand{\Ss}{S({\rm\halpha})_{\rm s}}
\newcommand{\EMs}{{\rm EM}_{\rm s}}
\newcommand{\EMsHalpha}{{\rm EM}(\halpha)_{\rm s}}
\newcommand{\EMsHbeta}{{\rm EM}(\hbeta)_{\rm s}}

\newcommand{\DMs}{\widehat{\rm DM}_{\rm s}}

\turnoffedit

\begin{document}

\received{\today}
\revised{\today}
\accepted{\today}
\submitjournal{ApJ}

\shorttitle{The host of FRB\,121102}
\shortauthors{Tendulkar et al.}

\title{The Host Galaxy and Redshift of the Repeating Fast Radio Burst FRB 121102}

\correspondingauthor{S.~P.~Tendulkar, C.~G.~Bassa}
\email{shriharsh@physics.mcgill.ca; bassa@astron.nl}

\author[0000-0003-2548-2926]{S.~P.~Tendulkar}
\affiliation{Department of Physics and McGill Space Institute, McGill University, 3600 University St., Montreal, QC H3A 2T8, Canada}

\author{C.~G.~Bassa}
\affiliation{ASTRON, the Netherlands Institute for Radio Astronomy, Postbus 2, NL-7990 AA Dwingeloo, The Netherlands}

\author{J.~M.~Cordes}
\affiliation{Cornell Center for Astrophysics and Planetary Science and Department of Astronomy, Cornell University, Ithaca, NY 14853, USA}

\author{G.~C.~Bower}
\affiliation{Academia Sinica Institute of Astronomy and Astrophysics, 645 N. A'ohoku Place, Hilo, HI 96720, USA}

\author{C.~J.~Law}
\affiliation{Department of Astronomy and Radio Astronomy Lab, University of California, Berkeley, CA 94720, USA}

\author{S.~Chatterjee}
\affiliation{Cornell Center for Astrophysics and Planetary Science and Department of Astronomy, Cornell University, Ithaca, NY 14853, USA}


\author{E.~A.~K.~Adams}
\affiliation{ASTRON, the Netherlands Institute for Radio Astronomy, Postbus 2, NL-7990 AA Dwingeloo, The Netherlands}

\author{S.~Bogdanov}
\affiliation{Columbia Astrophysics Laboratory, Columbia University,  New York, NY 10027, USA}

\author{S.~Burke-Spolaor}
\affiliation{National Radio Astronomy Observatory, Socorro, NM 87801, USA}
\affiliation{Department of Physics and Astronomy, West Virginia University, Morgantown, WV 26506, USA}
\affiliation{Center for Gravitational Waves and Cosmology, West Virginia University, Chestnut Ridge Research Building, Morgantown, WV 26505}

\author{B.~J.~Butler}
\affiliation{National Radio Astronomy Observatory, Socorro, NM 87801, USA}

\author{P.~Demorest}
\affiliation{National Radio Astronomy Observatory, Socorro, NM 87801, USA}

\author{J.~W.~T.~Hessels}
\affiliation{ASTRON, the Netherlands Institute for Radio Astronomy, Postbus 2, NL-7990 AA Dwingeloo, The Netherlands}
\affiliation{Anton Pannekoek Institute for Astronomy, University of Amsterdam, Science Park 904, 1098 XH Amsterdam, The Netherlands}

\author{V.~M.~Kaspi}
\affiliation{Department of Physics and McGill Space Institute, McGill University, 3600 University St., Montreal, QC H3A 2T8, Canada}

\author{T.~J.~W.~Lazio}
\affiliation{Jet Propulsion Laboratory, California Institute of Technology, Pasadena, CA 91109, USA}

\author{N.~Maddox}
\affiliation{ASTRON, the Netherlands Institute for Radio Astronomy, Postbus 2, NL-7990 AA Dwingeloo, The Netherlands}

\author{B.~Marcote}
\affiliation{Joint Institute for VLBI ERIC, Postbus 2, 7990 AA Dwingeloo, The Netherlands}

\author{M.~A.~McLaughlin}
\affiliation{Department of Physics and Astronomy, West Virginia University, Morgantown, WV 26506, USA}
\affiliation{Center for Gravitational Waves and Cosmology, West Virginia University, Chestnut Ridge Research Building, Morgantown, WV 26505}

\author{Z.~Paragi}
\affiliation{Joint Institute for VLBI ERIC, Postbus 2, 7990 AA Dwingeloo, The Netherlands}

\author{S.~M.~Ransom}
\affiliation{National Radio Astronomy Observatory, Charlottesville, VA 22903, USA}


\author{P.~Scholz}
\affiliation{National Research Council of Canada, Herzberg Astronomy and Astrophysics, Dominion Radio Astrophysical Observatory, P.O. Box 248, Penticton, BC V2A 6J9, Canada}

\author{A.~Seymour}
\affiliation{Arecibo Observatory, HC3 Box 53995, Arecibo, PR 00612, USA}

\author{L.~G.~Spitler}
\affiliation{Max-Planck-Institut f\"ur Radioastronomie, Auf dem H\"ugel 69, Bonn, D-53121, Germany}

\author{H.~J.~van Langevelde}
\affiliation{Joint Institute for VLBI ERIC, Postbus 2, 7990 AA Dwingeloo, The Netherlands}
\affiliation{Leiden Observatory, Leiden University, PO Box 9513, 2300 RA Leiden, The Netherlands}

\author{R.~S.~Wharton}
\affiliation{Cornell Center for Astrophysics and Planetary Science and Department of Astronomy, Cornell University, Ithaca, NY 14853, USA}

\begin{abstract}
The precise localization of the repeating fast radio burst (\frb) has provided the first unambiguous  association (chance coincidence probability $p\lesssim3\times10^{-4}$) of an FRB with an optical and persistent radio counterpart. We report on optical imaging and spectroscopy of the counterpart and find that it is an extended ($0\farcs6-0\farcs8$) object displaying prominent Balmer and [\ion{O}{3}] emission lines. 
Based on the spectrum and emission line ratios, we classify the counterpart as a low-metallicity, star-forming, $m_{r^\prime} = 25.1$\,AB mag dwarf galaxy at a redshift of $z=0.19273(8)$, corresponding to a luminosity distance of 972\,Mpc. From the angular size, the redshift, and luminosity, we estimate the host galaxy to have a diameter $\lesssim4$\,kpc and a stellar mass of $M_*\sim4-7\times 10^{7}\,M_\sun$, assuming a mass-to-light ratio between 2 to 3$\,M_\sun\,L_\sun^{-1}$. Based on the \halpha\ flux, we estimate the star formation rate of the host to be $0.4\,M_\sun\,\mathrm{yr^{-1}}$ and a substantial host dispersion measure depth $\lesssim 324\,\mathrm{pc\,cm^{-3}}$. The net dispersion measure contribution of the host galaxy to \frb\ is likely to be lower than this value depending on geometrical factors. We show that the persistent radio source at \frb's location reported by \citet{mph+16} is offset from the galaxy's center of light by $\sim$200\,mas and the host galaxy does not show optical signatures for AGN activity. If \frb\ is typical of the wider FRB population and if future interferometric localizations preferentially find them  in dwarf galaxies with low metallicities and prominent emission lines, they would share such a preference with long gamma ray bursts and superluminous supernovae. 


\end{abstract}


\section{Introduction}
 \label{sec:intro}
Fast radio bursts (FRBs) are bright ($\sim$Jy) and short ($\sim$ms) bursts of radio emission that have dispersion measures (DMs) in excess of the line of sight DM contribution expected from the electron distribution of our Galaxy. To date 18 FRBs have been reported
 --- most of them detected at the Parkes telescope \citep{lbm+07,tsb+13,bb14,kskl12,rsj15,pbb+15,kjb+16,cpk+16,rsb+16} and one each at the Arecibo \citep{sch+14} and Green Bank telescopes \citep{mls+15}. 

A plethora of source models have been proposed to explain the properties of FRBs \citep[see e.g.][for a brief review]{katz16}. According to the models, the excess DM for FRBs may be intrinsic to the source, placing it within the Galaxy; it may arise mostly from the intergalactic medium, placing a source of FRBs at cosmological distances ($z\sim0.2-1$) or it may arise from the host galaxy, placing a source of FRBs at extragalactic, but not necessarily cosmological, distances ($\sim100$\,Mpc).

Since the only evidence to claim an extragalactic origin for FRBs has been the anomalously high DM, some models also attempted to explain the excess DM as a part of the model, thus allowing FRBs to be Galactic. All FRBs observed to date have been detected with single dish radio telescopes, for which the localization is of order arcminutes, insufficient to obtain an unambiguous association with any object. To date, no independent information about their redshift, environment, and source could be obtained due to the lack of an accurate localization of FRBs. \citet{kjb+16} attempted to identify the host of FRB\,150418 on the basis of a fading radio source in the field that was localized to a $z=0.492$ galaxy. However, later work identified the radio source as a variable active galactic nucleus (AGN) that may not be related to the source \citep{wb16,bbt+16,gmg+16,jkb+16}.

 Repeated radio bursts were observed from the location of the Arecibo-detected \frb\ \citep{ssh+16a,ssh+16b}, with the same DM as the first detection, indicating a common source. As discussed by \citet{ssh+16a}, it is unclear whether the repetition makes \frb\ unique among known FRBs, or whether radio telescopes other than Arecibo lack the sensitivity to readily detect repeat bursts from other known FRBs.

\citet{clw+16} used the Karl G. Jansky Very Large Array (VLA) to directly localize the repeated bursts from \frb\ with 100-mas precision and reported an unresolved, persistent radio source and an extended optical counterpart at the location with a chance coincidence probability of $\approx 3\times10^{-4}$ --- the first unambiguous identification of multi-wavelength counterparts to FRBs. Independently, \citet{mph+16} used the European VLBI Network (EVN) to localize the bursts and the persistent source and showed that both are co-located within $\sim12$ milliarcseconds.

Here we report the imaging and spectroscopic follow-up of the optical counterpart to \frb\ using the 8-m Gemini North telescope. 




\section{Observations and Data Analysis}
\label{sec:obs}

The location of \frb\ was observed with the Gemini Multi-Object
Spectrograph (GMOS) instrument at the 8-m Gemini North telescope atop
Mauna Kea, Hawai'i. Imaging observations were obtained with SDSS
$r^\prime$, $i^\prime$ and $z^\prime$ filters on 2016 October 24, 25, and
November 2, under photometric and clear conditions with
$0\farcs58$ to $0\farcs66$ seeing. Exposure times of 250\,s were used in the $r^\prime$ filter and of 300\,s in the $i^\prime$ and $z^\prime$ filters with total exposures of 1250\,s in $r^\prime$, 1000\,s
in $i^\prime$ and 1500\,s in $z^\prime$. The detectors were read out with $2\times2$ binning, providing a pixel scale of $0\farcs146$\,pix$^{-1}$. The
images were corrected for a bias offset, as measured from the overscan
regions, flat fielded using sky flats and then registered and
co-added.

The images were astrometrically calibrated against the \textit{Gaia} DR1 Catalog \citep{bvp+16}. To limit the effects of distortion, the central $2\farcm2\times2\farcm2$ subsection of the images were used. Each of the $r^\prime$, $i^\prime$, and $z^\prime$ images were matched with 35 -- 50 unblended stars yielding an astrometric calibration with 7 -- 9\,mas root-mean-square (rms) position residuals in each coordinate after iteratively removing $\sim4-5$ outliers. The error in the mean astrometric position with respect to the \textit{Gaia} frame is thus $\sim1-2\,$mas.


We used the \texttt{Source Extractor} \citep{ba96} software to detect and extract sources in the coadded images. The $r^\prime$ and $i^\prime$ images were photometrically calibrated with respect to the IPHAS DR2 catalog \citep{bfd+14} using Vega-AB magnitude conversions stated therein. We measure isophotal integrated magnitudes of $m_{r^\prime} = 25.1\pm0.1$\,AB mag and $m_{i^\prime}=23.9\pm0.1$\,AB mag for the optical counterpart of \frb. The error value includes the photometric errors and rms zero-point scatter. Ongoing observations will provide full photometric calibration in $g^\prime$, $r^\prime$, $i^\prime$, and $z^\prime$ bands and will be reported in a subsequent publication. 

\begin{figure*}
    \includegraphics[width=\textwidth]{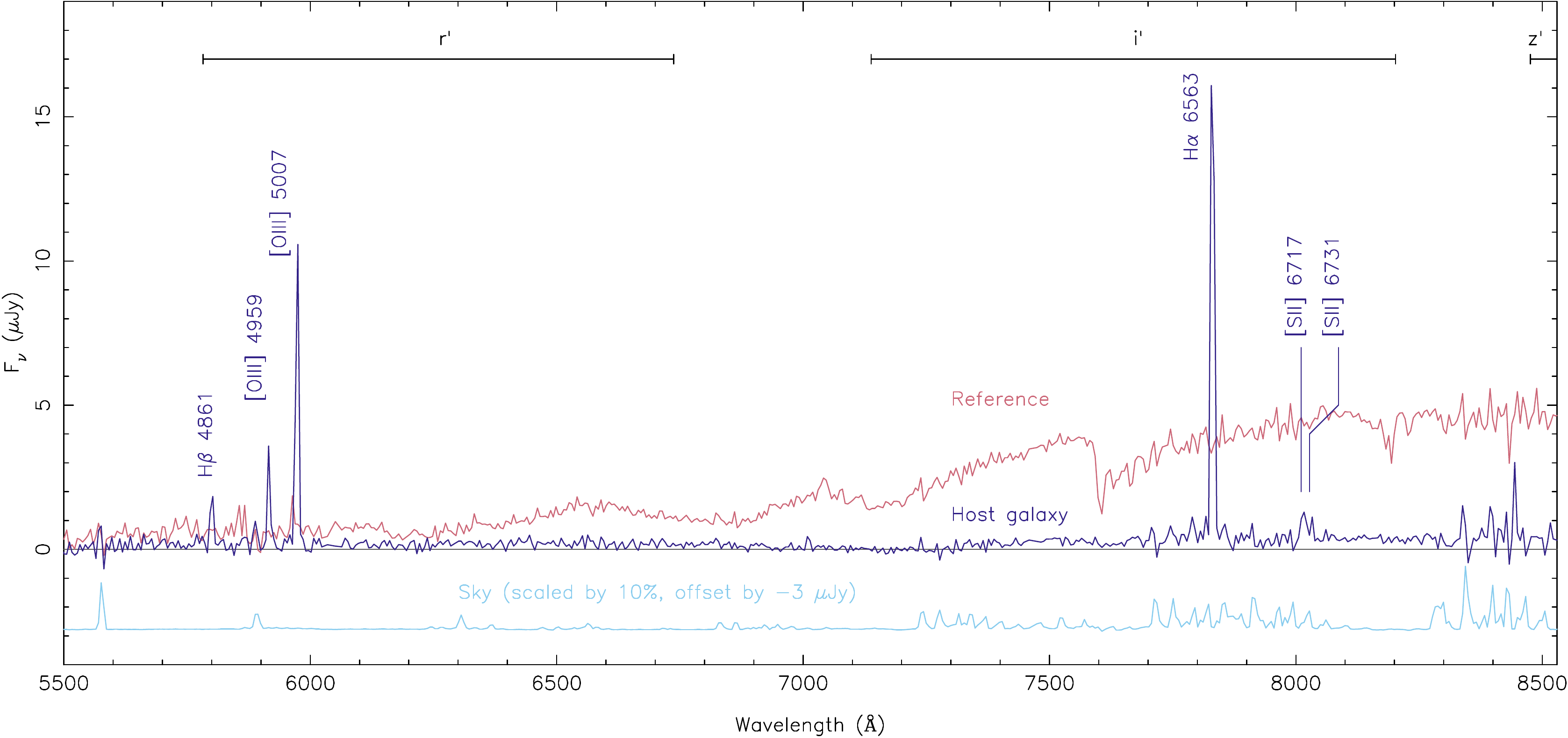}
  \caption{The co-added spectrum of the host galaxy of \frb,
    the reference object, and the sky contribution (scaled by 10\% and offset by $-3$\,$\upmu$Jy). The spectra
    have been resampled to the instrumental resolution. Prominent
    emission lines are labelled with their rest frame wavelengths. Black horizontal bars denote the wavelength ranges of the filters used for imaging. Most of the wavelength coverage of the $z^\prime$ band is outside the coverage of this plot.}
  \label{fig:spectrum}
\end{figure*}

Spectroscopic observations were obtained with GMOS on 2016 November 9
and 10 with the 400\,lines\,mm$^{-1}$ grating (R400) in combination
with a $1\arcsec$ slit, covering the wavelength range from 4650 to
8900\,\AA. A total of nine 1800\,s exposures were taken with
$2\times2$\,binning, providing a spatial scale of
$0\farcs292$\,pix$^{-1}$ and an instrumental resolution of 4.66\,\AA,
sampled at $1.36$\,\AA\,pix$^{-1}$. The conditions were clear, with
$0\farcs8$ to $1\farcs0$ seeing on the first night, and $0\farcs9$ to
$1\farcs1$ on the second. To aid the spectral extraction of the very faint counterpart, the slit was oriented at a position angle of
$18\fdg6$, containing the counterpart to \frb\ as well as
an $m_{r^\prime}=24.3$\, AB mag, $m_{i^\prime}=22.7$\,AB mag foreground star, located $2\farcs8$ to the South (shown later in Figure~\ref{fig:bands}). 

The low signal-to-noise of the spectral trace of the FRB counterpart
on the individual bias-corrected long-slit spectra complicated
spectral extraction through the optimal method by
\citet{hor86}. Instead, we used a variant of the optimal extraction method
of \citet{hyn02} by modelling the spectral trace of the reference
object by a Moffat function \citep{mof69} to determine the position
and width of the spatial profile as a function of wavelength. Because
of the proximity of the reference object to the FRB counterpart
(20\,pix), we assume that the spatial profile as a function of
wavelength is identical for both. We note that though the counterpart is slightly
resolved in the imaging observations, the worse seeing during the
spectroscopic observations (by a factor 1.2 to 1.9) means the seeing
dominates the spatial profile. The residual images validate this
assumption; no residual flux is seen once the extracted model is
subtracted from the image. To optimally extract the spectra of
the FRB counterpart, the reference object as well as the sky
background, we then simultaneously fit the spatial profile at the
location of the counterpart and at the location of the reference
object on top of a spatially varying linear polynomial for each column
in the dispersion direction.

Wavelength calibrations were obtained from arc lamp exposures,
modelling the dispersion location to wavelength through 4th order
polynomials, yielding rms residuals of better than 0.2\,\AA. The
individual wavelength calibrated spectra were then combined and
averaged. The instrumental response of the spectrograph was calibrated
using an observation of the spectrophotometric standard
\object{Hiltner\,600} \citep{hws+92,hsh+94}, which was taken on 2016
November 7 as part of the standard Gemini calibration plan with
identical instrumental setup as the science observations. The flux-calibrated spectrum of the reference object gives a spectroscopic AB magnitude of $m_{i^\prime}\approx22.6$, about 11\% higher than derived from photometry. Given that the spectrophotometric standard
was observed on a different night with worse seeing ($1\farcs4$), we
attribute this difference to slit losses and scale the flux of the observed
spectra of the reference object and the FRB counterpart by a factor
0.89.


\section{Results and Analysis}
\label{sec:results}


The final combined and calibrated spectrum is shown in Figure\,\ref{fig:spectrum}. Besides continuum emission, which is
weakly detected in the red part of the spectrum, four strong emission
lines are clearly visible and are identified as \halpha, \hbeta,
and [\ion{O}{3}] $\lambda4959$ and [\ion{O}{3}] $\lambda5007$ indicating that the optical counterpart is a star-forming galaxy. The corresponding weighted mean redshift is $z=0.19273\pm0.0008$. Weaker emission lines from [\ion{S}{2}] $\lambda\lambda6717$, 6731 are also
detected. The [\ion{N}{2}] $\lambda\lambda6549$, 6583 and the
[\ion{O}{1}] $\lambda6300$ lines are not seen.

Gaussian fits to the emission lines in the rest frame yield the flux
and 1-$\sigma$ width values listed in Table\,\ref{tab:lines}. We estimate rest frame equivalent widths for the strongest emission lines; $392\pm102$\,\AA\ for [\ion{O}{3}] $\lambda5007$ and $290\pm55$\,\AA\ for \halpha.

\begin{figure}
  \includegraphics[width=\columnwidth]{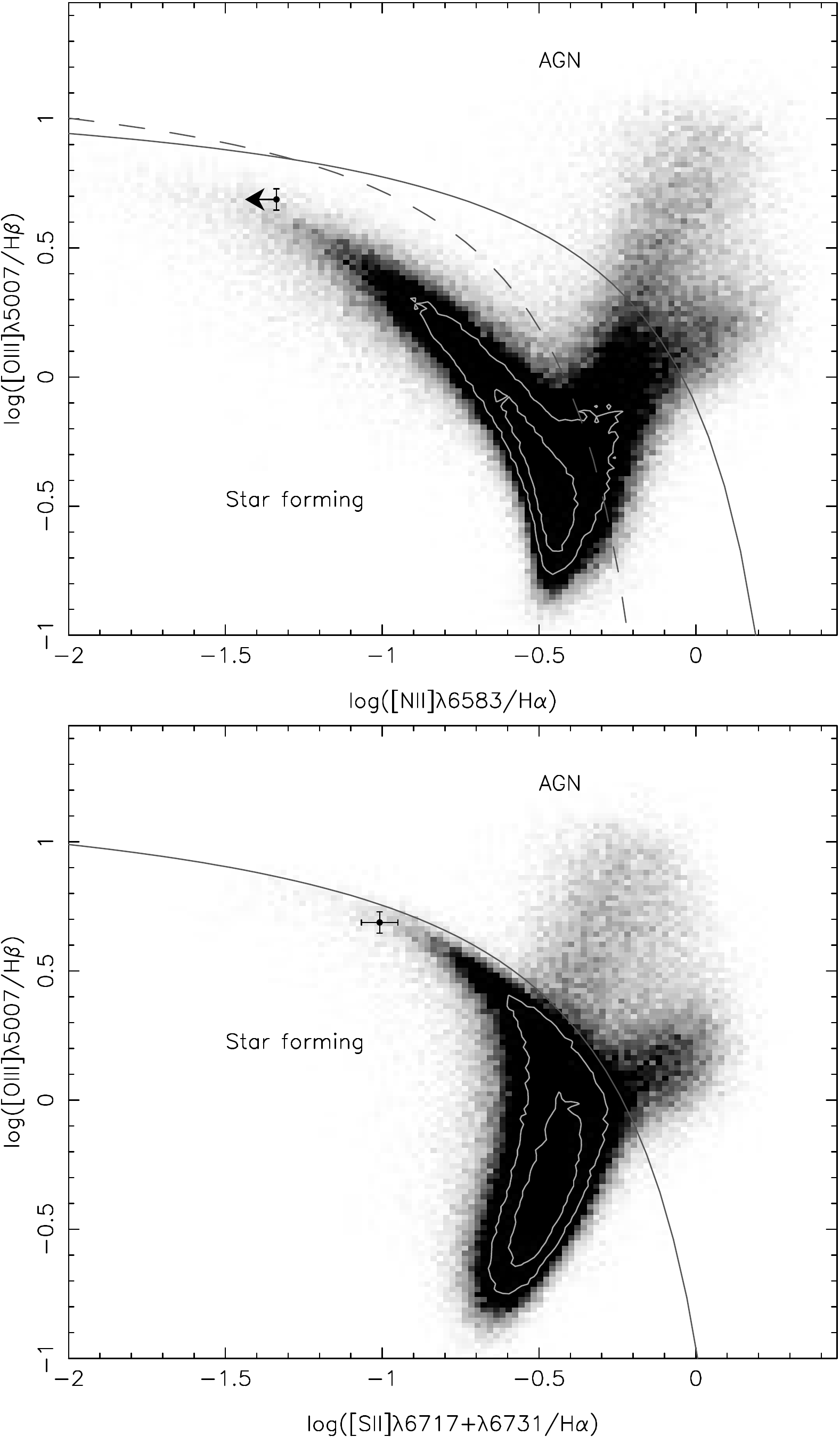}
  \caption{BPT \citep{bpt81} diagrams of [\ion{N}{2}]/\halpha\ and
    [\ion{S}{2}]/\halpha\ against [\ion{O}{3}]/\hbeta\ for the SDSS
    DR12 \citep{aaa+15} galaxy sample with significant ($>$5-$\sigma$)
    emission lines. The black symbol with error bars denotes the
    location of the host galaxy of FRB\,121102. The solid and dashed
    lines denote the demarcations between star-forming and AGN dominated
    galaxies, respectively \citep{kds+01,kgkh06,kht+03}. The region between the two curves corresponds to composite objects with AGN and star formation.}
  \label{fig:bpt}
\end{figure}

The ratios of measured line fluxes for [\ion{O}{3}]/\hbeta\ against
[\ion{N}{2}]/\halpha\ and [\ion{S}{2}]/\halpha --- the well-known Baldwin, Phillips \& Terlevich (BPT) diagram \citep{bpt81} --- are shown in Figure~\ref{fig:bpt}. The line ratios of the host galaxy of \frb\ are compared to those from the SDSS
DR12 galaxy sample \citep{aaa+15}. The locations below and to the left of the
solid and dashed grey lines indicate that the emission lines are due
to star formation and not due to AGN activity \citep{kds+01,kgkh06,kht+03}. Note that the BPT diagram line ratios are insensitive to reddening (from the Milky Way as well as the host itself).

\begin{deluxetable}{llll}
  \centering
  \tablecolumns{4}
  \tablecaption{Emission line properties.\label{tab:lines}}
  \tablewidth{0pt}
  \tabletypesize{\footnotesize}
  \tablehead{
    \colhead{Line} &
    \colhead{Obs. Flux} &
    \colhead{Width ($\sigma$)} &
    \colhead{$A_\lambda/A_V$}\\
    \colhead{} &
    \colhead{(erg\,cm$^{-2}$\,s$^{-1}$)} &
    \colhead{(\AA)} &
    \colhead{(mag)}
  }
  \startdata
  \hbeta\                       & $0.118(11)\times10^{-16}$ & 1.91(19) & 0.941 \\
  {[}\ion{O}{3}{]} $\lambda4959$ & $0.171(10)\times10^{-16}$ & 1.75(11) & 0.921 \\
  {[}\ion{O}{3}{]} $\lambda5007$ & $0.575(11)\times10^{-16}$ & 1.89(4)  & 0.911 \\
  {[}\ion{O}{1}{]} $\lambda6300$ & $<0.009\times10^{-16}$    &          & 0.670 \\
  {[}\ion{N}{2}{]} $\lambda6549$ & $<0.021\times10^{-16}$    &          & 0.625 \\
  \halpha\                      & $0.652(9)\times10^{-16}$  & 2.02(3)  & 0.622 \\
  {[}\ion{N}{2}{]} $\lambda6583$ & $<0.030\times10^{-16}$    &          & 0.619 \\
  {[}\ion{S}{2}{]} $\lambda6717$ & $0.040(6)\times10^{-16}$  & 2.4(4)   & 0.596 \\
  {[}\ion{S}{2}{]} $\lambda6731$ & $0.024(6)\times10^{-16}$  & 2.4(6)   & 0.593 \\
  \enddata
  \tablecomments{Observed emission line properties from fitting normalized Gaussians to the rest-wavelength host galaxy
    spectrum. Upper limits (3-$\sigma$) on line fluxes assume
    Gaussian widths of $\sigma=2$\,\AA. The absorption $A_\lambda/A_V$ at the
    observed line wavelengths is taken from \citet{ccm89}. To obtain
    unabsorbed line fluxes, multiply by $10^{0.4
      (A_\lambda/A_V)A_V}$, where $A_V$ is the Galactic absorption
    towards FRB\,121102.}
\end{deluxetable}

\begin{figure}
  \includegraphics[width=\columnwidth]{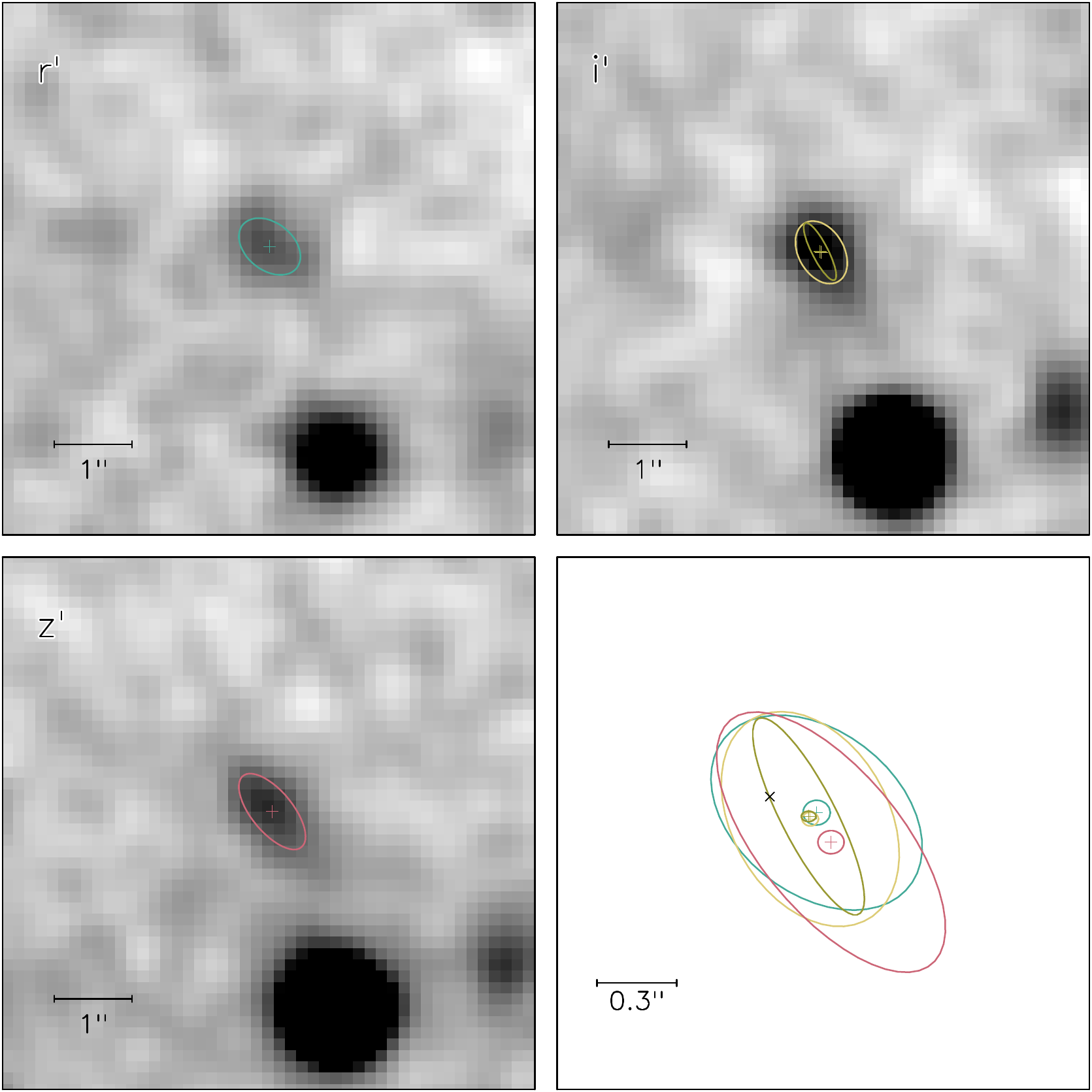}
  \caption{The top left, top right and bottom left panels show
    respective $7\farcs4\times7\farcs4$ subsections of the GMOS $r^\prime$,
    $i^\prime$ and $z^\prime$ images, centered on the optical counterpart to
    FRB\,121102. Each image has been smoothed by a Gaussian with a
    width of $0\farcs2$, while the plus sign and ellipse denote the
    position and extent of a two-dimensional Gaussian fit to the
    spatial profile of the counterpart. The $i^\prime$-band image also
    shows the narrower S\'ersic fit by \texttt{galfit}. The bottom
    right panel combines the positional and morphological measurements
    from the different bands on an astrometric frame of
    $1\arcsec\times1\arcsec$ in size. The colors are identical to
    those used in the other panels. The large ellipses denote the
    extent of the Gaussian and S\'ersic fits, while the small ellipses
    denote the 1-$\sigma$ absolute positional uncertainties. The
    location of the persistent counterpart as measured with the EVN at
    5\,GHz by \citet{mph+16} is represented by the black cross. The uncertainty in the EVN location is much smaller than the size of the symbol.)}
  \label{fig:bands}
\end{figure}

We use the \texttt{galfit} software \citep{phir02,phir10} to constrain
the morphology of the optical counterpart. A S\'ersic profile
($\Sigma(r)=\Sigma_\mathrm{e}e^{-\kappa[(r/R_\mathrm{e})^{1/n}-1]}$),
convolved with the point-spread-function, was fitted against the
spatial profile of the counterpart. For the $i^\prime$-band image, the
best fit has an effective radius of
$R_\mathrm{e}=0\farcs41\pm0\farcs06$, a S\'ersic index of
$n=2.2\pm1.5$, and an ellipticity of $b/a=0.25\pm0.13$. The lower
signal-to-noise of the counterpart in the $r^\prime$ and $z^\prime$
images did not permit meaningful results. Instead, we directly fit the
spatial profile in all three bands with a two-dimensional elliptical
Gaussian profile. In the case of the $i^\prime$-band image, the fit
provides a position and effective radius, taken as the Gaussian
$\sigma$, consistent with the S\'ersic profile convolved with the
point-spread-function. The results of the fits are shown in
Figure\,\ref{fig:bands}.

The position and extent of the host galaxy, as approximated with the
two-dimensional elliptical Gaussian profile, agrees well in the
$r^\prime$ and $i^\prime$ bands (semi-major axis $\sigma_a=0\farcs44$ with
ellipticity $b/a=0.68$), while the $z^\prime$-band has a
slightly offset position and appears larger ($\sigma=0\farcs59$ with
 $b/a=0.45$). We attribute this difference to the fact that the
the $r^\prime$ and $i^\prime$ bands are dominated by the bright
emission lines of \halpha, \hbeta, [\ion{O}{3}] $\lambda4959$ and
[\ion{O}{3}] $\lambda5007$, while the redder $z^\prime$-band traces
the continuum flux of the host galaxy. As such, the morphology
suggests that the host galaxy has at least one \ion{H}{2} region
at a slight offset from the galaxy center.


Finally, the bottom right panel of Figure\,\ref{fig:bands} plots the Gaussian centroids on the International Celestial Reference Frame (ICRF) through the astrometric calibration of the $r^\prime$, $i^\prime$, and $z^\prime$ images against \textit{Gaia}. The positional uncertainties in each axis are the quadratic sum of the astrometric tie against \textit{Gaia} (of order 2\,mas) and the centroid uncertainty on the image (between 20 and 50\,mas). The \textit{Gaia} frame is tied to the ICRF defined via radio VLBI to a $\sim$1\,mas precision \citep{mkl+16}, much smaller than the centroid uncertainty. We find that the position of the persistent radio source seen with the EVN at an observing frequency of 5\,GHz with a 1-mas precision \citep{mph+16}, is offset from the galaxy centroids by $186\pm68$ and $163\pm32$\,mas in the line-dominated $r^\prime$ and $i^\prime$ images, and $286\pm64$\,mas in the continuum-dominated $z^\prime$ image. Though offset from the centroids, the persistent radio source is located within the effective radii of the different bands.

\section{Discussion and Conclusions}
\label{sec:discussion}

The observations presented here confirm the interpretation by \citet{clw+16} that the extended optical counterpart associated with \frb\ is the host galaxy of the FRB. Our measurement of the redshift $z=0.19273$ is consistent with the DM-estimated value of $z_\mathrm{DM}<0.32$ \citep{clw+16} and together with the very low chance superposition probability, firmly places \frb\ at a cosmological distance, ruling out all Galactic models for this source. 


In the following discussion, we assume the cosmological parameters from the \citet{aaa+16} as implemented in \texttt{astropy.cosmology} \citep{astropy2013}, giving a luminosity distance of $D_\mathrm{L}=972$\,Mpc, and $1\arcsec$ corresponding to projected proper and comoving distances of 3.31\,kpc and 3.94\,kpc, respectively. 

We use the \citet{sfd98} estimate of the Galactic extinction along this line of sight\footnote{From the IRSA Dust Extinction Calculator \url{http://irsa.ipac.caltech.edu/applications/DUST/}}, $E_{B-V}=0.781$. Using $R_V=3.1$, we find $A_V=2.42$, and use the \citet{ccm89} Galactic extinction curve to correct the spectrum with band extinctions of $A_{r^\prime} = 2.15, A_{i^\prime} = 1.63,$ and $A_{z^\prime} = 1.16$\,mag. We note that the \citet{sfs10, sf11} recalibrated extinction model predicts a slightly lower extinction of  $E_{B-V}=0.673$. The results described below are insensitive to differences in the extinction at this level.  We do not apply $k$-correction to the magnitudes as they are not needed for the precision discussed here.

\subsection{Burst Energetics}
The redshift measurement allows us to put \frb's energetics on a firmer footing, confirming the energy scale of $10^{38}\,\mathrm{erg}\,(\delta\Omega/4\pi)(A_\nu/\mathrm{0.1\,Jy\,ms})(\Delta\nu/\mathrm{1\,GHz})$ calculated by \citet{clw+16} using a distance scale of 1\,Gpc. Here $A_\nu$ and $\Delta\nu$ are the fluence and bandwidth, respectively, at observing frequency $\nu$ and $\delta\Omega$ is the opening angle of the bursts. A more detailed analysis of energetics of individual bursts detected by the VLA and their rates will be reported in Law C. J. et al. (in preparation).

\subsection{Physical Properties of the Host}
The host of \frb\ is a small galaxy with a diameter of $\lesssim 4$\,kpc, inferred from the continuum-dominated $z^\prime$-band image. The absolute magnitudes, including the emission line fluxes and  after correcting for the Milky Way's extinction, are $M_{r^\prime} = -17.0$\,AB mag and $M_{i^\prime} = -17.7$\,AB mag, identifying the host as a dwarf galaxy. 

From Table~\ref{tab:lines}, the \halpha\ luminosity of the host galaxy, corrected for Milky Way extinction, is $L_\halpha = 2.9\times10^{40}\,\mathrm{erg\,s^{-1}}$. The corresponding star formation rate is $\SFR(\halpha) = 7.9\times10^{-42}\,M_\sun\,\mathrm{yr^{-1}}\times (L_\halpha/\mathrm{erg\,s^{-1}}) = 0.23\,M_\sun\,\mathrm{yr^{-1}}$ \citep{ktc1994}. This value does not completely account for the extinction of \halpha\ photons in the host galaxy. The correction suggested by \citet{kgjd2002} is $\SFR(IR) = 2.7\times \SFR(\halpha)^{1.3} \approx 0.4 \,M_\sun\,\mathrm{yr^{-1}}$ (in the 8--1000\,$\upmu$m band). This is consistent with the 3-$\sigma$ upper limit of $<9\,M_\sun\,\mathrm{yr^{-1}}$ estimated from the ALMA non-detection of the host at 230\,GHz assuming a submillimeter spectral index $\alpha=3$ \citep{clw+16}.

The mass-to-light ratio $\Upsilon_{*}$ is dependent on the star formation history and the initial mass function for star formation. As an estimate, we use $\Upsilon_{*}^R \approx 2-3\,M_\sun\,L_\sun^{-1}$ based on the dynamics of dwarf galaxies with high star formation rates \citep{lvf14}, implying a stellar mass $M_* \sim 4-7 \times 10^7\ M_\sun$. As dwarf galaxies are usually gas-rich \citep[e.g. ][]{pch+12}, we expect that this estimate is a lower limit to the host baryonic mass. We also note that dwarf galaxies are typically dark matter dominated \citep{ccf00}, and so the total dynamical mass is likely to be larger.

We use the $R_{23}$ \citep{kd02}, $N2$, $O3N2$ \citep{pp04}, and the recently defined diagnostic of \citet[][labelled here as $y$]{dksn16} to estimate the metallicity where,
\be
R_{23} &=& \log_{10}((\text{[\ion{O}{2}]}\lambda3727+\text{[\ion{O}{3}]}\lambda\lambda4959,5007)/\hbeta),\nonumber\\
N2 &=& \log_{10}(\text{[\ion{N}{2}]}\lambda6584/\halpha),\nonumber\\
O3N2 &=& \log_{10}(\text{[\ion{O}{3}]}\lambda5007)/\text{[\ion{N}{2}]}\lambda6584 \times \halpha/\hbeta), \text{and}\nonumber \\ 
y &=&  \log_{10}(\text{[\ion{N}{2}]}\lambda6584/\text{[\ion{S}{2}]}\lambda\lambda6717,6731) \nonumber\\
& &+0.264 \log_{10}(\text{[\ion{N}{2}]}\lambda6584/\halpha). \nonumber
\ee

As the [\ion{O}{2}]$\lambda$3727 line is outside our spectral coverage and [\ion{N}{2}]$\lambda6584$ is not detected, we can only set an upper limit to the metallicity. Using the extinction-corrected line fluxes, we measure, 
\be 
R_{23}& \geq & 0.77, \nonumber\\
N2 & \leq & -1.34,\nonumber \\
O3N2 & \geq & 2.1, \nonumber \\
y & \leq & -0.66, \nonumber 
\ee
where the limits are calculated from the 3-$\sigma$ limit on [\ion{N}{2}]$\lambda$6584 flux and assuming the lower limit for the unmeasured  [\ion{O}{2}]$\lambda$3727 flux to be zero. This corresponds to a 3-$\sigma$ metallicity limit of $\log_{10}([\mathrm{O/H}]) + 12 < 8.4$ \citep{kd02}, $<8.4$ \citep[][, $N2$]{pp04}, $<8.4$ \citep[][$O3N2$]{pp04}\footnote{We note that the \citet{pp04} calibration has high scatter for $O3N2\gtrsim2$ but the limit quoted here includes the scatter.} and $<8.1$ \citep[][not including scatter]{dksn16}. We convert these into the oft-used KK04 scale \citep{kk04} using the conversions of \citet{ke08}. All measurements are consistent with $\log_{10}([\mathrm{O/H}]) + 12 \lesssim 8.7$ in the KK04 scale. The metallicity of the host is low  --- less than $\sim$15\% of all galaxies brighter than $M_B<-16$ have metallicity lower than 8.7 \citep[][]{gf15b}. This set of galaxies account for less than $20$\% of the star formation of the local Universe.

The host properties are similar to those of extreme emission line galaxies  \citep[EELGs; ][]{ass+11}, young, low-mass starbursts which have emission lines of rest-frame equivalent widths greater than 200\AA. 

\subsection{Ionized Gas  Properties in the Host}


The Balmer lines from the host also allow us to estimate the properties its ionized ISM  and its contribution to the total DM of \frb.

The $\Halpha$ surface density for the galaxy with  flux $F_{\halpha}$, semi-major axis $a$, and semi-minor axis $b$ is
\be
S(\halpha) &=& \frac{F_{\halpha}}{\pi a b}, \nonumber\\
&\approx&  6.8 \times 10^{-16}\,\mathrm{erg\,cm^{-2}\,s^{-1}\,arcsecond^{-2}}, \nonumber\\
&\approx&  120 \,\mathrm{Rayleigh},
\ee
where we have used the extinction corrected flux $F_{\halpha} = 2.6\times10^{-16}\,\mathrm{erg\,cm^{-2}\,s^{-1}}$ and the semi-major and minor axes ($a = 0\farcs44$, $b/a=0.68$) from the $i^\prime$ and $r^\prime$ images. In the source frame (denoted below by the subscript, `s'), the surface density is
\be
\Ss = (1+z)^4 S(\halpha) = 243\,\mathrm{Rayleigh}. 
\ee

For a temperature  $T = 10^4 T_4$~K, we express the  emission measure ($\EM\ = \int n_e^2 \mathrm{d}s$) given by \citet{reyn77}  in the galaxy's frame
\be
\EMsHalpha &=& 2.75 \mathrm{pc\,cm^{-6}} \, T_4^{0.9}\,  \left[\frac{\Ss}{\mathrm{Rayleigh}}\right], \nonumber \\
&\approx& 670\,\mathrm{pc\,cm^{-6}}\,T_4^{0.9}.
\ee
We get a smaller  value from the extinction-corrected  $\hbeta$ flux,  $\EMsHbeta \approx 530\,\mathrm{pc\,cm^{-6}}$. For the calculations below, we proceed with a combined  estimate,  $\EMs\ \approx 600\,\mathrm{pc\,cm^{-6}}$.

This value is fairly large compared to measurements of the local Galactic disk. The WHAM  $\Halpha$ survey, for example, gives values of tens of pc~cm$^{-6}$ in the Galactic plane  and about 1 pc~cm$^{-6}$ looking out of the plane \citep[][]{hbk+08}. However, 
lines of sight to distant pulsars and studies of other galaxies give EM values in the hundreds \citep[][]{reyn77, hdb+09}. 

The estimate for $\EMs$ is sensitive to the inferred solid angle of the galaxy and emitting regions. Ongoing observations with the \textit{Hubble Space Telescope} will better resolve the $\halpha$ emitting structures and improve our constraint on the EM with respect to the location of the burst. 

 The implied optical depth for free-free absorption at an observation frequency $\nu$ (in GHz) is
 \be
\tau_{\rm ff} &\approx&  3.3\times10^{-6} [(1+z)\nu_\mathrm{GHz}]^{-2.1} T_4^{-1.35} \EMs \nonumber \\ 
&\approx& 1.4 \times 10^{-3}  \nu_\mathrm{GHz}^{-2.1} T_4^{-0.45}.
\ee
 Free-free absorption for \frb\ is therefore negligible even at 100\,MHz. This suggests that the radio spectra of the bursts and possibly the persistent source  are unaffected by absorption and are inherent to the emission process or to propagation effects near the sources, confirming the inference made by \citep{ssh+16b} based on the widely varying spectral shapes of the bursts alone.

\subsubsection{Implied DM from $\Halpha$-emitting Gas}
The EM implies a DM value sometimes given by $\DM = (\EM f_{\rm f} L)^{1/2}$, where
$f_{\rm f}$ is the volume filling factor of ionized clouds in a region of total size $L$ \citep{reyn77}. As summarized in Appendix~B of \citet[][]{cws+16},  additional fluctuations  decrease the  DM derived from EM, giving a source-frame value,
\be
\DMs &\approx & 
387\ \text{pc cm$^{-3}$}\, 
	L_{\rm kpc}^{1/2}
	\left[ \frac{f_{\rm f}}{\zeta(1+\epsilon^2)/4}\right]^{1/2} \nonumber \\
& & \times \left(\frac{\EM}{600\ \text{pc cm$^{-6}$}} \right)^{1/2},
\label{eq:em2dm}
\ee
 where $\epsilon\le 1$ is the fractional variation inside discrete clouds due to turbulent-like density variations and  $\zeta\ge 1$ defines cloud-to-cloud density variations  in the ionized region of depth  $L_{\rm kpc}$  in kpc. Here we have used $\EMs\ = 600$ \text{pc cm$^{-6}$} and assumed 100\% cloud-to-cloud variations ($\zeta = 2$) and fully modulated electron densities inside clouds ($\epsilon=1$).


The host contribution to the {\it measured} \DM\ is a factor $(1+z)^{-1}$ smaller than the source frame DM\footnote{The factor of $(1+z)^{-1}$ is a combination of the photon redshift, time dilation and the frequency$^{-2}$ dependence of cold plasma dispersion.}.
Also, the line of sight to the  FRB source may  sample only a fraction of  $\DMs$  depending on if
it is embedded in or offset from the $\Halpha$-emitting gas.  For an effective path length through the ionized gas
$L_{\rm FRB} \le L$,  we then have
%
\be
\DMhatFRB &=& \frac{\DMs}{1+z} \left(\frac{L_{\rm FRB}}{L} \right) 
\nonumber \\
&&\hspace{-0.85in}   \approx  324 \ \text{pc cm$^{-3}$}  \left(\frac{L_{\rm FRB}}{L} \right)
\left[\frac{4L_{\rm kpc}f_{\rm f}}{ \zeta(1+\epsilon^2)} \right]^{1/2}.
\ee

This estimate can be compared with empirical constraints  discussed in  \citet{clw+16} on contributions from the host and the intergalactic medium (IGM) to the total \DM\  made by subtracting the NE2001 model's DM contribution from the Milky Way \citep{cl02} 
($\DM_{\rm MW} = 188 \ \text{pc cm$^{-3}$}$) and the Milky Way halo ($\DM_{\rm MW_{halo}} = 30 \ \text{pc cm$^{-3}$}$) from the total $\DM\ = 558\,\mathrm{pc\,cm^{-3}}$. This gives $\DM_\mathrm{IGM} + \DM_\mathrm{host} = 340\,\mathrm{pc\,cm^{-3}}$.  
The Milky Way contributions have uncertain errors but are likely of order 20\%. 
The  measured redshift implies a mean IGM contribution $\DM_{\rm IGM} \approx 200\ \text{pc cm$^{-3}$}$ \citep{ioka03,inou04} but can vary by about $\pm 85 \ \text{pc cm$^{-3}$}$ \citep[][]{mcqu14}.   This yields a range of possible values for $\DM_{\rm host}$: $55 \lesssim \DM_{\rm host} \lesssim 225\,\mathrm{pc\,cm^{-3}}$ that further implies   
$0.09 \lesssim (L_{\rm FRB} / L) \left [L_{\rm kpc}  f_{\rm f} /  \zeta(1+\epsilon^2)\right]^{1/2} \lesssim 0.35$.
The ionized region therefore must have some degree of clumpiness or the effective path length is significantly  smaller than the size of the ionized region. 

Radio pulsars  in the Large and Small Magellanic Clouds have DMs spanning the range 45--273\,$\mathrm{pc\,cm^{-3}}$
 and  70--200 $\mathrm{pc\,cm^{-3}}$, respectively \citep{mhth05}. 
This empirically demonstrates that the free electron content of star-forming dwarf galaxies is of the order we estimate. The relatively large DM contribution from the host galaxy (as inferred from the $\Halpha$ emission) implies that any contributions from the vicinity of the FRB source itself are probably quite small. This may rule out a very young ($<100$\,yr) supernova remnant \citep[e.g. ][]{piro16}. 

\subsection{Implications for Source Models}
\citet{clw+16} reported the locations of the radio bursts, the optical and variable radio counterparts and the absence of millimeter-wave and X-ray emission. \citet{mph+16} have shown that the bursts and the persistent radio source are colocated to within a linear projected separation of 40\,pc, suggesting that the two emission sources should be physically related, though not necessarily the same source. The radio source properties are consistent with a low luminosity AGN or a young ($<$1000\,yr) supernova remnant (SNR) powered by an energetic neutron star \citep[e.g. ][]{mkm16}. 

The optical properties of the galaxies reported here do not add support to the AGN interpretation although it cannot be conclusively ruled out. The BPT diagnostics for the host (Figure~\ref{fig:bpt}) show no indication of AGN activity. However, this may not be conclusive as the majority of radio-loud AGN show no optical signatures of activity \citep{ms07}. This is further supported by five low luminosity AGN with no optical signatures have also recently been discovered \citep{pyop16}. However, these objects are almost exclusively hosted in galaxies with much larger stellar masses ($\sim10^{10}\,M_\sun$).
We also note that the radio source is offset from the optical center of the galaxy by 170--300\,mas, corresponding to a transverse linear distance of 0.5--1\,kpc, nearly a quarter to half of the radial extent, which is not consistent with a central AGN, but such offsets have been seen before in dwarf galaxies, e.g. Henize 2-10 \citep{rsjb11}. 

The association of an optical/X-ray AGN with a dwarf galaxy is also extremely rare. A search of emission-line dwarf galaxies ($10^{8.5} \lesssim M_* \lesssim 10^{9.5}\,M_\sun$) using BPT line diagnostics identified an AGN rate of $\sim$ 0.5 \% \citep{rgg13}, with an additional 0.05 \% of dwarf galaxies searched exhibiting narrow emission lines consistent with star formation band broad \halpha\ consistent with an AGN. Similarly, an X-ray survey of $z<1$ dwarf galaxies reported an AGN rate of 0.6--3\% \citep{pgg+16}. Of the dwarf galaxies known to host AGN, only two exhibit nuclear radio emission that appears to originate from a black-hole jet, Henize 2-10 and Mrk 709 \citep{rsjb11,rpr+14}.  Both have strong nuclear X-ray emission that originates from the AGN but optical emission lines that are dominated by star-formation processes. The combination of a compact radio source, absent nuclear X-ray emission, strong star-formation optical emission lines, and weak or non-existent broad optical emission lines that we observe in the host of \frb\ has no analogue in any known galaxy to the best of our knowledge. 

The high star formation rate is consistent with the presence of a young SNR or a cluster of young massive stars (i.e. an OB association), which would naturally link FRBs to neutron stars which are the favored progenitor models.

\subsubsection{Relation to Dwarf Galaxies}
It is interesting to note that the only FRB host directly identified so far is a low metallicity dwarf galaxy rather than, say, an extremely high-star-formation-rate galaxy such as Arp\,220 or a galaxy with a very powerful AGN or some other extreme characteristics. Dwarf galaxies are also a small fraction of the stellar mass in the Universe \citep{pch+12}. \citet{rsb+16} also suggested that the extremely low scattering of FRB\,150807 compared to its DM may be linked to its origin from a low-mass ($<10^{9}\,M_\sun$) galaxy. However, the strong polarization and scattering properties of FRB\,110523 do suggest the presence of turbulent magnetized plasma around the source \citep{mls+15}, suggesting that individual FRB environments may be quite diverse.


If FRBs are indeed more commonly hosted by dwarf galaxies in the low redshift Universe, they would share this preference with two other classes of high-energy transients --- long duration gamma-ray bursts and superluminous supernovae, both of which prefer low-mass, low-metallicity, and high star formation rate hosts \citep[e.g., ][ and other works]{fls+06,plt+13,vsj+15,pqy+16}. Indeed, superluminous supernovae are prefentially hosted by EELGs \citep{lsk+15}. If this relation is true, it may point to a link between FRBs and extremely massive progenitor stars, possibly extending to magnetars that have been associated with massive progenitor stars \citep[e.g.  ][]{ok14}.

\subsection{Future Optical Follow-Up of FRBs}
A link between FRBs and dwarf galaxies will impact future multi-wavelength follow-up plans. Without the precise localization for \frb\ \citep{clw+16}, the host galaxy is scarcely distinguishable from other objects in the deep Gemini images. 

 Due to the trade-off between field of view and localization precision, FRB search projects that have a large FRB detection rate such as CHIME (Kaspi V. M. et al,. 2017, in preparation), UTMOST \citep{cfb+16}, and HIRAX \citep{nbb+16} will localize high signal to noise detections to only sub-arcmin precision. If FRB hosts are star-forming galaxies with strong emission lines, slitless objective prism spectroscopy could efficiently distinguish these objects from a field of stars and elliptical galaxies, leading to putative host identifications without very precise localization. However, this strongly depends on the link between FRBs and their host properties and the homogeneity of FRBs --- which will first have to be confirmed with more interferometric localizations.


We note, of course, that our above discussion regarding the possible relationship between FRBs and dwarf galaxies in general is based on a single data point of a repeating FRB, which may not be representative of the broader FRB population \citep[see][for more details]{ssh+16a,ssh+16b}.

\acknowledgements 
We are very grateful to the staff of the Gemini Observatory for their help and flexibility throughout this program. We also thank R.~F.~Trainor and A.~Delahaye for helpful discussions.

Our work is based on observations obtained at the Gemini Observatory (program GN-2016B-DD-2), which is operated by the Association of Universities for Research in Astronomy, Inc., under a cooperative agreement with the NSF on behalf of the Gemini partnership: the National Science Foundation (United States), the National Research Council (Canada), CONICYT (Chile), Ministerio de Ciencia, Tecnolog\'{i}a e Innovaci\'{o}n Productiva (Argentina), and Minist\'{e}rio da Ci\^{e}ncia, Tecnologia e Inova\c{c}\~{a}o (Brazil). 

This work has made use of data from the European Space Agency (ESA) mission {\it Gaia} (\url{http://www.cosmos.esa.int/gaia}), processed by the {\it Gaia} Data Processing and Analysis Consortium (DPAC, \url{http://www.cosmos.esa.int/web/gaia/dpac/consortium}). Funding for the DPAC has been provided by national institutions, in particular the institutions participating in the {\it Gaia} Multilateral Agreement. This research made use of Astropy, a community-developed core Python package for Astronomy (Astropy Collaboration, 2013, \url{http://www.astropy.org}).

S.P.T acknowledges support from a McGill Astrophysics postdoctoral fellowship. The research leading to these results has received funding from the European Research Council (ERC) under the European Union's Seventh Framework Programme (FP7/2007-2013). C.G.B. and J.W.T.H. gratefully acknowledge funding for this work from ERC Starting Grant DRAGNET under contract number 337062. J.M.C., R.S.W., and S.C. acknowledge prior support from the National Science Foundation through grants AST-1104617 and AST-1008213. This work was partially supported by the University of California Lab Fees program under award number LF-12-237863. The research leading to these results has received funding from the European Research Council (ERC) under the European Union’s Seventh Framework Programme (FP7/2007-2013).  J.W.T.H. is an NWO Vidi Fellow. V.M.K. holds the Lorne Trottier and a Canada Research Chair and receives support from an NSERC Discovery Grant and Accelerator Supplement, from a R. Howard Webster Foundation Fellowship from the Canadian Institute for Advanced Research (CIFAR), and from the FRQNT Centre de Recherche en Astrophysique du Quebec. B.M. acknowledges support by the Spanish Ministerio de Econom\'ia y Competitividad (MINECO/FEDER, UE) under grants AYA2013-47447-C3-1-P, AYA2016-76012-C3-1-P, and MDM-2014-0369 of ICCUB (Unidad de Excelencia `Mar\'ia de Maeztu'). L.G.S. gratefully acknowledge financial support from the ERC Starting Grant BEACON under contract number 279702 and the Max Planck Society. Part of this research was carried out at the Jet Propulsion Laboratory, California Institute of Technology, under a contract with the National Aeronautics and Space Administration. E.A.K.A. is supported by TOP1EW.14.105, which is financed by the Netherlands Organisation for Scientific Research (NWO). M.A.M. is supported by NSF award \#1458952. S.B.S is a Jansky Fellow of the National Radio Astronomy Observatory. P.S. is a Covington Fellow at the Dominion Radio Astrophysical Observatory.

\facility{Gemini:Gillett (GMOS)}
\software{ESO-MIDAS, astro-py, galfit, SExtractor}

\begin{thebibliography}{}
\expandafter\ifx\csname natexlab\endcsname\relax\def\natexlab#1{#1}\fi
\providecommand{\url}[1]{\href{#1}{#1}}

\bibitem[{{Alam} {et~al.}(2015){Alam}, {Albareti}, {Allende Prieto}, {Anders},
  {Anderson}, {Anderton}, {Andrews}, {Armengaud}, {Aubourg}, {Bailey}, \&
  et~al.}]{aaa+15}
{Alam}, S., {Albareti}, F.~D., {Allende Prieto}, C., {et~al.} 2015, \apjs, 219,
  12

\bibitem[{{Astropy Collaboration} {et~al.}(2013){Astropy Collaboration},
  {Robitaille}, {Tollerud}, {Greenfield}, {Droettboom}, {Bray}, {Aldcroft},
  {Davis}, {Ginsburg}, {Price-Whelan}, {Kerzendorf}, {Conley}, {Crighton},
  {Barbary}, {Muna}, {Ferguson}, {Grollier}, {Parikh}, {Nair}, {Unther},
  {Deil}, {Woillez}, {Conseil}, {Kramer}, {Turner}, {Singer}, {Fox}, {Weaver},
  {Zabalza}, {Edwards}, {Azalee Bostroem}, {Burke}, {Casey}, {Crawford},
  {Dencheva}, {Ely}, {Jenness}, {Labrie}, {Lim}, {Pierfederici}, {Pontzen},
  {Ptak}, {Refsdal}, {Servillat}, \& {Streicher}}]{astropy2013}
{Astropy Collaboration}, {Robitaille}, T.~P., {Tollerud}, E.~J., {et~al.} 2013,
  \aap, 558, A33

\bibitem[{{Atek} {et~al.}(2011){Atek}, {Siana}, {Scarlata}, {Malkan},
  {McCarthy}, {Teplitz}, {Henry}, {Colbert}, {Bridge}, {Bunker}, {Dressler},
  {Fosbury}, {Hathi}, {Martin}, {Ross}, \& {Shim}}]{ass+11}
{Atek}, H., {Siana}, B., {Scarlata}, C., {et~al.} 2011, \apj, 743, 121

\bibitem[{{Baldwin} {et~al.}(1981){Baldwin}, {Phillips}, \&
  {Terlevich}}]{bpt81}
{Baldwin}, J.~A., {Phillips}, M.~M., \& {Terlevich}, R. 1981, \pasp, 93, 5


\bibitem[{{Barentsen} {et~al.}(2014){Barentsen}, {Farnhill}, {Drew},
  {Gonz{\'a}lez-Solares}, {Greimel}, {Irwin}, {Miszalski}, {Ruhland}, {Groot},
  {Mampaso}, {Sale}, {Henden}, {Aungwerojwit}, {Barlow}, {Carter}, {Corradi},
  {Drake}, {Eisl{\"o}ffel}, {Fabregat}, {G{\"a}nsicke}, {Gentile Fusillo},
  {Greiss}, {Hales}, {Hodgkin}, {Huckvale}, {Irwin}, {King}, {Knigge},
  {Kupfer}, {Lagadec}, {Lennon}, {Lewis}, {Mohr-Smith}, {Morris}, {Naylor},
  {Parker}, {Phillipps}, {Pyrzas}, {Raddi}, {Roelofs}, {Rodr{\'{\i}}guez-Gil},
  {Sabin}, {Scaringi}, {Steeghs}, {Suso}, {Tata}, {Unruh}, {van Roestel},
  {Viironen}, {Vink}, {Walton}, {Wright}, \& {Zijlstra}}]{bfd+14}
{Barentsen}, G., {Farnhill}, H.~J., {Drew}, J.~E., {et~al.} 2014, \mnras, 444,
  3230

\bibitem[{{Bassa} {et~al.}(2016){Bassa}, {Beswick}, {Tingay}, {Keane},
  {Bhandari}, {Johnston}, {Totani}, {Tominaga}, {Yasuda}, {Stappers}, {Barr},
  {Kramer}, \& {Possenti}}]{bbt+16}
{Bassa}, C.~G., {Beswick}, R., {Tingay}, S.~J., {et~al.} 2016, \mnras, 463, L36

\bibitem[{{Bertin} \& {Arnouts}(1996)}]{ba96}
{Bertin}, E., \& {Arnouts}, S. 1996, \aaps, 117, 393

\bibitem[{{Burke-Spolaor} \& {Bannister}(2014)}]{bb14}
{Burke-Spolaor}, S., \& {Bannister}, K.~W. 2014, \apj, 792, 19

\bibitem[{{Caleb} {et~al.}(2016){Caleb}, {Flynn}, {Bailes}, {Barr}, {Bateman},
  {Bhandari}, {Campbell-Wilson}, {Green}, {Hunstead}, {Jameson}, {Jankowski},
  {Keane}, {Ravi}, {van Straten}, \& {Krishnan}}]{cfb+16}
{Caleb}, M., {Flynn}, C., {Bailes}, M., {et~al.} 2016, \mnras, 458, 718

\bibitem[{{Cardelli} {et~al.}(1989){Cardelli}, {Clayton}, \& {Mathis}}]{ccm89}
{Cardelli}, J.~A., {Clayton}, G.~C., \& {Mathis}, J.~S. 1989, \apj, 345, 245

\bibitem[{{Champion} {et~al.}(2016){Champion}, {Petroff}, {Kramer}, {Keith},
  {Bailes}, {Barr}, {Bates}, {Bhat}, {Burgay}, {Burke-Spolaor}, {Flynn},
  {Jameson}, {Johnston}, {Ng}, {Levin}, {Possenti}, {Stappers}, {van Straten},
  {Thornton}, {Tiburzi}, \& {Lyne}}]{cpk+16}
{Champion}, D.~J., {Petroff}, E., {Kramer}, M., {et~al.} 2016, \mnras, 460, L30

\bibitem[{{Chatterjee} {et~al.}(2017){Chatterjee}, {Law}, {Wharton},
  {Burke-Spolaor}, {Hessels}, {Bower}, {Cordes}, {Tendulkar}, {Bassa},
  {Demorest}, {Butler}, {Seymour}, {Scholz}, {Abruzzo}, {Bogdanov}, {Kaspi},
  {Keimpema}, {Lazio}, {Marcote}, {McLaughlin}, {Paragi}, {Ransom}, {Rupen},
  {Spitler}, \& {van Langevelde}}]{clw+16}
{Chatterjee}, S., {Law}, C.~J., {Wharton}, R.~S., {et~al.} 2017, \nat, 000, 000

\bibitem[{{Cordes} \& {Lazio}(2002)}]{cl02}
{Cordes}, J.~M., \& {Lazio}, T.~J.~W. 2002, ArXiv Astrophysics e-prints,
  astro-ph/0207156

\bibitem[{{Cordes} {et~al.}(2016){Cordes}, {Wharton}, {Spitler}, {Chatterjee},
  \& {Wasserman}}]{cws+16}
{Cordes}, J.~M., {Wharton}, R.~S., {Spitler}, L.~G., {Chatterjee}, S., \&
  {Wasserman}, I. 2016, ArXiv e-prints, arXiv:1605.05890

\bibitem[{{C{\^o}t{\'e}} {et~al.}(2000){C{\^o}t{\'e}}, {Carignan}, \&
  {Freeman}}]{ccf00}
{C{\^o}t{\'e}}, S., {Carignan}, C., \& {Freeman}, K.~C. 2000, \aj, 120, 3027

\bibitem[{{Dopita} {et~al.}(2016){Dopita}, {Kewley}, {Sutherland}, \&
  {Nicholls}}]{dksn16}
{Dopita}, M.~A., {Kewley}, L.~J., {Sutherland}, R.~S., \& {Nicholls}, D.~C.
  2016, \apss, 361, 61

\bibitem[{{Fruchter} {et~al.}(2006){Fruchter}, {Levan}, {Strolger},
  {Vreeswijk}, {Thorsett}, {Bersier}, {Burud}, {Castro Cer{\'o}n},
  {Castro-Tirado}, {Conselice}, {Dahlen}, {Ferguson}, {Fynbo}, {Garnavich},
  {Gibbons}, {Gorosabel}, {Gull}, {Hjorth}, {Holland}, {Kouveliotou}, {Levay},
  {Livio}, {Metzger}, {Nugent}, {Petro}, {Pian}, {Rhoads}, {Riess}, {Sahu},
  {Smette}, {Tanvir}, {Wijers}, \& {Woosley}}]{fls+06}
{Fruchter}, A.~S., {Levan}, A.~J., {Strolger}, L., {et~al.} 2006, \nat, 441,
  463

\bibitem[{{Gaia Collaboration} {et~al.}(2016){Gaia Collaboration}, {Brown},
  {Vallenari}, {Prusti}, {de Bruijne}, {Mignard}, {Drimmel}, \&
  {co-authors}}]{bvp+16}
{Gaia Collaboration}, {Brown}, A.~G.~A., {Vallenari}, A., {et~al.} 2016, ArXiv
  e-prints

\bibitem[{{Giroletti} {et~al.}(2016){Giroletti}, {Marcote}, {Garrett},
  {Paragi}, {Yang}, {Hada}, {Muxlow}, \& {Cheung}}]{gmg+16}
{Giroletti}, M., {Marcote}, B., {Garrett}, M.~A., {et~al.} 2016, \aap, 593, L16

\bibitem[{{Graham} \& {Fruchter}(2015)}]{gf15b}
{Graham}, J.~F., \& {Fruchter}, A.~S. 2015, ArXiv e-prints, arXiv:1511.01079

\bibitem[{{Haffner} {et~al.}(2009){Haffner}, {Dettmar}, {Beckman}, {Wood},
  {Slavin}, {Giammanco}, {Madsen}, {Zurita}, \& {Reynolds}}]{hdb+09}
{Haffner}, L.~M., {Dettmar}, R.-J., {Beckman}, J.~E., {et~al.} 2009, Reviews of
  Modern Physics, 81, 969

\bibitem[{{Hamuy} {et~al.}(1994){Hamuy}, {Suntzeff}, {Heathcote}, {Walker},
  {Gigoux}, \& {Phillips}}]{hsh+94}
{Hamuy}, M., {Suntzeff}, N.~B., {Heathcote}, S.~R., {et~al.} 1994, \pasp, 106,
  566

\bibitem[{{Hamuy} {et~al.}(1992){Hamuy}, {Walker}, {Suntzeff}, {Gigoux},
  {Heathcote}, \& {Phillips}}]{hws+92}
{Hamuy}, M., {Walker}, A.~R., {Suntzeff}, N.~B., {et~al.} 1992, \pasp, 104, 533

\bibitem[{{Hill} {et~al.}(2008){Hill}, {Benjamin}, {Kowal}, {Reynolds},
  {Haffner}, \& {Lazarian}}]{hbk+08}
{Hill}, A.~S., {Benjamin}, R.~A., {Kowal}, G., {et~al.} 2008, \apj, 686, 363

\bibitem[{{Horne}(1986)}]{hor86}
{Horne}, K. 1986, \pasp, 98, 609

\bibitem[{{Hynes}(2002)}]{hyn02}
{Hynes}, R.~I. 2002, \aap, 382, 752

\bibitem[{{Inoue}(2004)}]{inou04}
{Inoue}, S. 2004, \mnras, 348, 999

\bibitem[{{Ioka}(2003)}]{ioka03}
{Ioka}, K. 2003, \apjl, 598, L79

\bibitem[{{Johnston} {et~al.}(2017){Johnston}, {Keane}, {Bhandari}, {Macquart},
  {Tingay}, {Barr}, {Bassa}, {Beswick}, {Burgay}, {Chandra}, {Honma}, {Kramer},
  {Petroff}, {Possenti}, {Stappers}, \& {Sugai}}]{jkb+16}
{Johnston}, S., {Keane}, E.~F., {Bhandari}, S., {et~al.} 2017, \mnras, 465,
  2143

\bibitem[{{Katz}(2016)}]{katz16}
{Katz}, J.~I. 2016, Modern Physics Letters A, 31, 1630013

\bibitem[{{Kauffmann} {et~al.}(2003){Kauffmann}, {Heckman}, {Tremonti},
  {Brinchmann}, {Charlot}, {White}, {Ridgway}, {Brinkmann}, {Fukugita}, {Hall},
  {Ivezi{\'c}}, {Richards}, \& {Schneider}}]{kht+03}
{Kauffmann}, G., {Heckman}, T.~M., {Tremonti}, C., {et~al.} 2003, \mnras, 346,
  1055

\bibitem[{{Keane} {et~al.}(2012){Keane}, {Stappers}, {Kramer}, \&
  {Lyne}}]{kskl12}
{Keane}, E.~F., {Stappers}, B.~W., {Kramer}, M., \& {Lyne}, A.~G. 2012, \mnras,
  425, L71

\bibitem[{{Keane} {et~al.}(2016){Keane}, {Johnston}, {Bhandari}, {Barr},
  {Bhat}, {Burgay}, {Caleb}, {Flynn}, {Jameson}, {Kramer}, {Petroff},
  {Possenti}, {van Straten}, {Bailes}, {Burke-Spolaor}, {Eatough}, {Stappers},
  {Totani}, {Honma}, {Furusawa}, {Hattori}, {Morokuma}, {Niino}, {Sugai},
  {Terai}, {Tominaga}, {Yamasaki}, {Yasuda}, {Allen}, {Cooke}, {Jencson},
  {Kasliwal}, {Kaplan}, {Tingay}, {Williams}, {Wayth}, {Chandra}, {Perrodin},
  {Berezina}, {Mickaliger}, \& {Bassa}}]{kjb+16}
{Keane}, E.~F., {Johnston}, S., {Bhandari}, S., {et~al.} 2016, \nat, 530, 453

\bibitem[{{Kennicutt} {et~al.}(1994){Kennicutt}, {Tamblyn}, \&
  {Congdon}}]{ktc1994}
{Kennicutt}, Jr., R.~C., {Tamblyn}, P., \& {Congdon}, C.~E. 1994, \apj, 435, 22

\bibitem[{{Kewley} \& {Dopita}(2002)}]{kd02}
{Kewley}, L.~J., \& {Dopita}, M.~A. 2002, \apjs, 142, 35

\bibitem[{{Kewley} {et~al.}(2001){Kewley}, {Dopita}, {Sutherland}, {Heisler},
  \& {Trevena}}]{kds+01}
{Kewley}, L.~J., {Dopita}, M.~A., {Sutherland}, R.~S., {Heisler}, C.~A., \&
  {Trevena}, J. 2001, \apj, 556, 121

\bibitem[{{Kewley} \& {Ellison}(2008)}]{ke08}
{Kewley}, L.~J., \& {Ellison}, S.~L. 2008, \apj, 681, 1183

\bibitem[{{Kewley} {et~al.}(2002){Kewley}, {Geller}, {Jansen}, \&
  {Dopita}}]{kgjd2002}
{Kewley}, L.~J., {Geller}, M.~J., {Jansen}, R.~A., \& {Dopita}, M.~A. 2002,
  \aj, 124, 3135

\bibitem[{{Kewley} {et~al.}(2006){Kewley}, {Groves}, {Kauffmann}, \&
  {Heckman}}]{kgkh06}
{Kewley}, L.~J., {Groves}, B., {Kauffmann}, G., \& {Heckman}, T. 2006, \mnras,
  372, 961

\bibitem[{{Kobulnicky} \& {Kewley}(2004)}]{kk04}
{Kobulnicky}, H.~A., \& {Kewley}, L.~J. 2004, \apj, 617, 240

\bibitem[{{Lelli} {et~al.}(2014){Lelli}, {Verheijen}, \& {Fraternali}}]{lvf14}
{Lelli}, F., {Verheijen}, M., \& {Fraternali}, F. 2014, \aap, 566, A71

\bibitem[{{Leloudas} {et~al.}(2015){Leloudas}, {Schulze}, {Kr{\"u}hler},
  {Gorosabel}, {Christensen}, {Mehner}, {de Ugarte Postigo}, {Amor{\'{\i}}n},
  {Th{\"o}ne}, {Anderson}, {Bauer}, {Gallazzi}, {He{\l}miniak}, {Hjorth},
  {Ibar}, {Malesani}, {Morell}, {Vinko}, \& {Wheeler}}]{lsk+15}
{Leloudas}, G., {Schulze}, S., {Kr{\"u}hler}, T., {et~al.} 2015, \mnras, 449,
  917

\bibitem[{{Lorimer} {et~al.}(2007){Lorimer}, {Bailes}, {McLaughlin},
  {Narkevic}, \& {Crawford}}]{lbm+07}
{Lorimer}, D.~R., {Bailes}, M., {McLaughlin}, M.~A., {Narkevic}, D.~J., \&
  {Crawford}, F. 2007, Science, 318, 777

\bibitem[{{Manchester} {et~al.}(2005){Manchester}, {Hobbs}, {Teoh}, \&
  {Hobbs}}]{mhth05}
{Manchester}, R.~N., {Hobbs}, G.~B., {Teoh}, A., \& {Hobbs}, M. 2005, \aj, 129,
  1993

\bibitem[{{Marcote} {et~al.}(2017){Marcote}, {Paragi}, {Hessels}, {Keimpema},
  \& {van Langevelde}}]{mph+16}
{Marcote}, B., {Paragi}, Z., {Hessels}, J.~W.~T., {Keimpema}, A., \& {van
  Langevelde}, H.~J. e.~a. 2017, \apj, 000, 000

\bibitem[{{Masui} {et~al.}(2015){Masui}, {Lin}, {Sievers}, {Anderson}, {Chang},
  {Chen}, {Ganguly}, {Jarvis}, {Kuo}, {Li}, {Liao}, {McLaughlin}, {Pen},
  {Peterson}, {Roman}, {Timbie}, {Voytek}, \& {Yadav}}]{mls+15}
{Masui}, K., {Lin}, H.-H., {Sievers}, J., {et~al.} 2015, \nat, 528, 523

\bibitem[{{Mauch} \& {Sadler}(2007)}]{ms07}
{Mauch}, T., \& {Sadler}, E.~M. 2007, \mnras, 375, 931

\bibitem[{{McQuinn}(2014)}]{mcqu14}
{McQuinn}, M. 2014, \apjl, 780, L33

\bibitem[{{Mignard} {et~al.}(2016){Mignard}, {Klioner}, {Lindegren}, {Bastian},
  {Bombrun}, {Hernandez}, {Hobbs}, {Lammers}, {Michalik}, {Ramos-Lerate},
  {Biermann}, {Butkevich}, {Comoretto}, {Joliet}, {Holl}, {Hutton}, {Parsons},
  {Steidelmueller}, {Andrei}, {Bourda}, \& {Charlot}}]{mkl+16}
{Mignard}, F., {Klioner}, S., {Lindegren}, L., {et~al.} 2016, ArXiv e-prints,
  arXiv:1609.07255

\bibitem[{{Moffat}(1969)}]{mof69}
{Moffat}, A.~F.~J. 1969, \aap, 3, 455

\bibitem[{{Murase} {et~al.}(2016){Murase}, {Kashiyama}, \&
  {M{\'e}sz{\'a}ros}}]{mkm16}
{Murase}, K., {Kashiyama}, K., \& {M{\'e}sz{\'a}ros}, P. 2016, \mnras, 461,
  1498

\bibitem[{{Newburgh} {et~al.}(2016){Newburgh}, {Bandura}, {Bucher}, {Chang},
  {Chiang}, {Cliche}, {Dav{\'e}}, {Dobbs}, {Clarkson}, {Ganga}, {Gogo},
  {Gumba}, {Gupta}, {Hilton}, {Johnstone}, {Karastergiou}, {Kunz}, {Lokhorst},
  {Maartens}, {Macpherson}, {Mdlalose}, {Moodley}, {Ngwenya}, {Parra},
  {Peterson}, {Recnik}, {Saliwanchik}, {Santos}, {Sievers}, {Smirnov},
  {Stronkhorst}, {Taylor}, {Vanderlinde}, {Van Vuuren}, {Weltman}, \&
  {Witzemann}}]{nbb+16}
{Newburgh}, L.~B., {Bandura}, K., {Bucher}, M.~A., {et~al.} 2016, in \procspie,
  Vol. 9906, Society of Photo-Optical Instrumentation Engineers (SPIE)
  Conference Series, 99065X

\bibitem[{{Olausen} \& {Kaspi}(2014)}]{ok14}
{Olausen}, S.~A., \& {Kaspi}, V.~M. 2014, \apjs, 212, 6

\bibitem[{{Papastergis} {et~al.}(2012){Papastergis}, {Cattaneo}, {Huang},
  {Giovanelli}, \& {Haynes}}]{pch+12}
{Papastergis}, E., {Cattaneo}, A., {Huang}, S., {Giovanelli}, R., \& {Haynes},
  M.~P. 2012, \apj, 759, 138

\bibitem[{{Pardo} {et~al.}(2016){Pardo}, {Goulding}, {Greene}, {Somerville},
  {Gallo}, {Hickox}, {Miller}, {Reines}, \& {Silverman}}]{pgg+16}
{Pardo}, K., {Goulding}, A.~D., {Greene}, J.~E., {et~al.} 2016, \apj, 831, 203

\bibitem[{{Park} {et~al.}(2016){Park}, {Yang}, {Oonk}, \& {Paragi}}]{pyop16}
{Park}, S., {Yang}, J., {Oonk}, J.~B.~R., \& {Paragi}, Z. 2016, ArXiv e-prints,
  arXiv:1611.05986

\bibitem[{{Peng} {et~al.}(2002){Peng}, {Ho}, {Impey}, \& {Rix}}]{phir02}
{Peng}, C.~Y., {Ho}, L.~C., {Impey}, C.~D., \& {Rix}, H.-W. 2002, \aj, 124, 266

\bibitem[{{Peng} {et~al.}(2010){Peng}, {Ho}, {Impey}, \& {Rix}}]{phir10}
---. 2010, \aj, 139, 2097

\bibitem[{{Perley} {et~al.}(2013){Perley}, {Levan}, {Tanvir}, {Cenko}, {Bloom},
  {Hjorth}, {Kr{\"u}hler}, {Filippenko}, {Fruchter}, {Fynbo}, {Jakobsson},
  {Kalirai}, {Milvang-Jensen}, {Morgan}, {Prochaska}, \& {Silverman}}]{plt+13}
{Perley}, D.~A., {Levan}, A.~J., {Tanvir}, N.~R., {et~al.} 2013, \apj, 778, 128

\bibitem[{{Perley} {et~al.}(2016){Perley}, {Quimby}, {Yan}, {Vreeswijk}, {De
  Cia}, {Lunnan}, {Gal-Yam}, {Yaron}, {Filippenko}, {Graham}, {Laher}, \&
  {Nugent}}]{pqy+16}
{Perley}, D.~A., {Quimby}, R.~M., {Yan}, L., {et~al.} 2016, \apj, 830, 13

\bibitem[{{Petroff} {et~al.}(2015){Petroff}, {Bailes}, {Barr}, {Barsdell},
  {Bhat}, {Bian}, {Burke-Spolaor}, {Caleb}, {Champion}, {Chandra}, {Da Costa},
  {Delvaux}, {Flynn}, {Gehrels}, {Greiner}, {Jameson}, {Johnston}, {Kasliwal},
  {Keane}, {Keller}, {Kocz}, {Kramer}, {Leloudas}, {Malesani}, {Mulchaey},
  {Ng}, {Ofek}, {Perley}, {Possenti}, {Schmidt}, {Shen}, {Stappers},
  {Tisserand}, {van Straten}, \& {Wolf}}]{pbb+15}
{Petroff}, E., {Bailes}, M., {Barr}, E.~D., {et~al.} 2015, \mnras, 447, 246

\bibitem[{{Pettini} \& {Pagel}(2004)}]{pp04}
{Pettini}, M., \& {Pagel}, B.~E.~J. 2004, \mnras, 348, L59

\bibitem[{{Piro}(2016)}]{piro16}
{Piro}, A.~L. 2016, \apjl, 824, L32

\bibitem[{{Planck Collaboration} {et~al.}(2016){Planck Collaboration}, {Ade},
  {Aghanim}, {Arnaud}, {Ashdown}, {Aumont}, {Baccigalupi}, {Banday},
  {Barreiro}, {Bartlett}, \& et~al.}]{aaa+16}
{Planck Collaboration}, {Ade}, P.~A.~R., {Aghanim}, N., {et~al.} 2016, \aap,
  594, A13

\bibitem[{{Ravi} {et~al.}(2015){Ravi}, {Shannon}, \& {Jameson}}]{rsj15}
{Ravi}, V., {Shannon}, R.~M., \& {Jameson}, A. 2015, \apjl, 799, L5

\bibitem[{{Ravi} {et~al.}(2016){Ravi}, {Shannon}, {Bailes}, {Bannister},
  {Bhandari}, {Bhat}, {Burke-Spolaor}, {Caleb}, {Flynn}, {Jameson}, {Johnston},
  {Keane}, {Kerr}, {Tiburzi}, {Tuntsov}, \& {Vedantham}}]{rsb+16}
{Ravi}, V., {Shannon}, R.~M., {Bailes}, M., {et~al.} 2016, ArXiv e-prints,
  arXiv:1611.05758

\bibitem[{{Reines} {et~al.}(2013){Reines}, {Greene}, \& {Geha}}]{rgg13}
{Reines}, A.~E., {Greene}, J.~E., \& {Geha}, M. 2013, \apj, 775, 116

\bibitem[{{Reines} {et~al.}(2014){Reines}, {Plotkin}, {Russell}, {Mezcua},
  {Condon}, {Sivakoff}, \& {Johnson}}]{rpr+14}
{Reines}, A.~E., {Plotkin}, R.~M., {Russell}, T.~D., {et~al.} 2014, \apjl, 787,
  L30

\bibitem[{{Reines} {et~al.}(2011){Reines}, {Sivakoff}, {Johnson}, \&
  {Brogan}}]{rsjb11}
{Reines}, A.~E., {Sivakoff}, G.~R., {Johnson}, K.~E., \& {Brogan}, C.~L. 2011,
  \nat, 470, 66

\bibitem[{{Reynolds}(1977)}]{reyn77}
{Reynolds}, R.~J. 1977, \apj, 216, 433

\bibitem[{{Schlafly} \& {Finkbeiner}(2011)}]{sf11}
{Schlafly}, E.~F., \& {Finkbeiner}, D.~P. 2011, \apj, 737, 103

\bibitem[{{Schlafly} {et~al.}(2010){Schlafly}, {Finkbeiner}, {Schlegel},
  {Juri{\'c}}, {Ivezi{\'c}}, {Gibson}, {Knapp}, \& {Weaver}}]{sfs10}
{Schlafly}, E.~F., {Finkbeiner}, D.~P., {Schlegel}, D.~J., {et~al.} 2010, \apj,
  725, 1175

\bibitem[{{Schlegel} {et~al.}(1998){Schlegel}, {Finkbeiner}, \&
  {Davis}}]{sfd98}
{Schlegel}, D.~J., {Finkbeiner}, D.~P., \& {Davis}, M. 1998, \apj, 500, 525

\bibitem[{{Scholz} {et~al.}(2016){Scholz}, {Spitler}, {Hessels}, {Chatterjee},
  {Cordes}, {Kaspi}, {Wharton}, {Bassa}, {Bogdanov}, {Camilo}, {Crawford},
  {Deneva}, {van Leeuwen}, {Lynch}, {Madsen}, {McLaughlin}, {Mickaliger},
  {Parent}, {Patel}, {Ransom}, {Seymour}, {Stairs}, {Stappers}, \&
  {Tendulkar}}]{ssh+16b}
{Scholz}, P., {Spitler}, L.~G., {Hessels}, J.~W.~T., {et~al.} 2016, \apj, 883,
  177

\bibitem[{{Spitler} {et~al.}(2014){Spitler}, {Cordes}, {Hessels}, {Lorimer},
  {McLaughlin}, {Chatterjee}, {Crawford}, {Deneva}, {Kaspi}, {Wharton},
  {Allen}, {Bogdanov}, {Brazier}, {Camilo}, {Freire}, {Jenet},
  {Karako-Argaman}, {Knispel}, {Lazarus}, {Lee}, {van Leeuwen}, {Lynch},
  {Ransom}, {Scholz}, {Siemens}, {Stairs}, {Stovall}, {Swiggum},
  {Venkataraman}, {Zhu}, {Aulbert}, \& {Fehrmann}}]{sch+14}
{Spitler}, L.~G., {Cordes}, J.~M., {Hessels}, J.~W.~T., {et~al.} 2014, \apj,
  790, 101

\bibitem[{{Spitler} {et~al.}(2016){Spitler}, {Scholz}, {Hessels}, {Bogdanov},
  {Brazier}, {Camilo}, {Chatterjee}, {Cordes}, {Crawford}, {Deneva}, {Ferdman},
  {Freire}, {Kaspi}, {Lazarus}, {Lynch}, {Madsen}, {McLaughlin}, {Patel},
  {Ransom}, {Seymour}, {Stairs}, {Stappers}, {van Leeuwen}, \& {Zhu}}]{ssh+16a}
{Spitler}, L.~G., {Scholz}, P., {Hessels}, J.~W.~T., {et~al.} 2016, \nat, 531,
  202

\bibitem[{{Thornton} {et~al.}(2013){Thornton}, {Stappers}, {Bailes},
  {Barsdell}, {Bates}, {Bhat}, {Burgay}, {Burke-Spolaor}, {Champion}, {Coster},
  {D'Amico}, {Jameson}, {Johnston}, {Keith}, {Kramer}, {Levin}, {Milia}, {Ng},
  {Possenti}, \& {van Straten}}]{tsb+13}
{Thornton}, D., {Stappers}, B., {Bailes}, M., {et~al.} 2013, Science, 341, 53

\bibitem[{{Vergani} {et~al.}(2015){Vergani}, {Salvaterra}, {Japelj}, {Le
  Floc'h}, {D'Avanzo}, {Fernandez-Soto}, {Kr{\"u}hler}, {Melandri}, {Boissier},
  {Covino}, {Puech}, {Greiner}, {Hunt}, {Perley}, {Petitjean}, {Vinci},
  {Hammer}, {Levan}, {Mannucci}, {Campana}, {Flores}, {Gomboc}, \&
  {Tagliaferri}}]{vsj+15}
{Vergani}, S.~D., {Salvaterra}, R., {Japelj}, J., {et~al.} 2015, \aap, 581,
  A102

\bibitem[{{Williams} \& {Berger}(2016)}]{wb16}
{Williams}, P.~K.~G., \& {Berger}, E. 2016, \apjl, 821, L22

\end{thebibliography}

\end{document}